\documentclass[twoside]{article}
\usepackage{epsf,latexsym,amssymb %,helvet,pifont
}
\makeatletter
\@addtoreset{equation}{section}

\makeatother

\usepackage{fancyheadings}
\renewcommand{\sectionmark}[1]%
                {\markboth{\thesection\ #1}{\thesection\ #1}}
\renewcommand{\subsectionmark}[1]%
                {\markright{\thesection\ #1}}

\makeatletter
\long\def\@makecaption#1#2{
 \vskip 10pt
 \setbox\@tempboxa\hbox{{\small\bf#1:} \small\sl#2}
 \ifdim \wd\@tempboxa >\hsize {\small\bf#1:} \small\sl#2\par
 \else \hbox to\hsize{\hfil\box\@tempboxa\hfil}
 \fi}
\makeatother

\begin{document}

%%%%%%%%%%%%%%%%%%%%%%%%% titlepage %%%%%%%%%%%%%%%%%%%%%%%%%%%%%%%%%
\thispagestyle{empty}
\renewcommand{\thefootnote}{\fnsymbol{footnote}}
\parskip0pt
%\begin{center}
%\fbox{\fbox{\bf Institut f\"ur Theoretische Physik der 
%Universit\"at Regensburg}}
%\end{center}

\vspace{0.5cm}
\hrule
\vspace{1pt}
\hrule
\vspace{5pt}
\hspace{0.5em} March 27, 1996 \hfill TPR--95--31, hep-th/9604015
\vspace{5pt}
\hrule
\vspace{1pt}
\hrule

\vspace{2cm}
\begin{center}
{\bf \large RADIAL PROPAGATORS AND WILSON LOOPS}

\vspace{1cm}
Stefan Leupold\footnote{stefan.leupold@physik.uni-regensburg.de}

\vspace{0.3cm}
Institut f\"ur Theoretische Physik, Universit\"at Regensburg,

D-93040 Regensburg, Germany

\vspace{0.8cm}
Heribert Weigert\footnote{weigert@mnhepw.hep.umn.edu}

\vspace{0.3cm}
School of Physics and Astronomy, University of Minnesota,

Minneapolis, MN 55455, USA
\end{center}

\vspace{2cm}

%%%%%%%%%%%%%%%%%%%%%%%%% abstract %%%%%%%%%%%%%%%%%%%%%%%%%%%%%%%%%%%
\begin{abstract}
We present a relation which
connects the propagator in the radial (Fock-Schwinger) gauge with a
gauge invariant Wilson loop. It is closely related to the well-known
field strength formula and can be used to calculate the radial gauge
propagator. The result is shown to diverge in four-dimensional space
even for free fields, its singular nature is however naturally
explained using the renormalization properties of Wilson loops with
cusps and self-intersections. Using this observation we provide a
consistent regularization scheme to facilitate loop calculations.
Finally we compare our results with previous approaches to derive a
propagator in Fock-Schwinger gauge.
\end{abstract}

\vspace{1cm}

%% resetting things after titlepage
\clearpage
\pagenumbering{arabic}
\renewcommand{\thefootnote}{\arabic{footnote}}
\parskip0.3cm
%%%%%%%%%%%%%%%%%%%%%%%%% end Titlepage %%%%%%%%%%%%%%%%%%%%%%%%%%%%%%

%%%%%%%%%%%%%%%%%%%%%%%%% Introduction %%%%%%%%%%%%%%%%%%%%%%%%%%%%%%%
\section{Introduction}

While perturbation theory for gauge fields formulated in covariant
gauges is very well established \cite{pascual} many aspects of
non-covariant gauges are still under discussion. In principle one
expects physical quantities to be independent of the chosen gauge.
However this might lead to the naive conclusion that a quantum theory
in an arbitrary gauge is simply obtained by inserting the respective
gauge fixing term and the appropriate Faddeev-Popov ghosts in the path
integral representation and reading off the Feynman rules.
Unfortunately it is not so easy to obtain the correct Feynman rules,
i.e.~a set of rules yielding the same results for observable
quantities as calculations in covariant gauges. Prominent examples are
formulations in temporal and axial gauges. Such gauge choices are
considered since one expects the Faddeev-Popov ghosts to decouple.
However problems even start with the determination of the appropriate
free gauge propagators. Temporal and axial gauge choices yield
propagators plagued by gauge poles in their momentum space
representation.  These are caused by the fact that such gauge
conditions are insufficient to {\it completely\/} remove the gauge
degrees of freedom. The correct treatment of such poles can cause
ghost fields to reappear \cite{cheng}, can break translational
invariance \cite{cara} or both \cite{leroy}. While these problems seem
to be ``restricted'' to the evaluation of the correct gauge
propagators and ghost fields, the necessity of introducing even new
multi-gluon vertices appears in the Coulomb gauge \cite{christ}. These
additional vertices are due to operator ordering problems which are
difficult to handle in the familiar path integral approach. They give
rise to anomalous interaction terms at the two-loop level \cite{doust}
and cause still unsolved problems with renormalization at the
three-loop level \cite{taylor}.

In this article we are interested in the radial (Fock-Schwinger) gauge
condition
\begin{equation}
x_\mu A^\mu(x) = 0 \,. \label{eq:fsgaugecond}
\end{equation}
It found widespread use in the context of QCD sum-rules (e.g.
\cite{shif}). There it is used as being more or less synonymous to the
important field strength formula
\begin{equation}
  \label{eq:fsformula}
  A_\mu^{\mbox{\scriptsize rad}}(x) = \int\limits^1_0 \!\! ds \, s x^\nu
  F_{\nu\mu}(sx)
\end{equation}
which enormously simplifies the task of organizing the operator product 
expansion of QCD n-point functions in terms of gauge invariant quantities
by expressing the gauge potential via the gauge covariant
field strength tensor.  It was introduced long ago \cite{fock}, \cite{schwing}
and rediscovered several times (e.g.~\cite{cron}). 

Only a few efforts have been made to establish perturbation theory for
radial gauge. The main reason for this is that the gauge condition
breaks translational invariance since the origin (in general an
arbitrary but fixed point $z$, c.f. (\ref{eq:arbz})) is singled out by
the gauge condition. Thus perturbation theory cannot be formulated in
momentum space as usual but must be set up in coordinate space.

The first attempt to evaluate the free radial propagator was performed
in \cite{kumm}. Later however the function $\Gamma_{\mu\nu}(x,y)$
presented there was shown to be not symmetric \cite{moda}. Moreover it
could not be symmetrized by adding $\Gamma_{\nu\mu}(y,x)$ since the
latter is not a solution of the free Dyson equation.  It was even
suspected in \cite{moda} that it might be impossible to find a
symmetric solution of this equation in four-dimensional space, due to
the appearance of divergences even on the level of the {\it free}
propagator when one uses the field strength formula to derive a free
propagator. Indeed we agree with this statement in principle, but we
will present an explanation for this problem and a way to bypass it.
Other approaches to define a radial gauge propagator try to circumvent
the problem (e.g. \cite{menot}) by sacrificing the field strength
formula as given in (\ref{eq:fsformula}) which was one of the main
reasons the gauge became popular in non-perturbative QCD sum rule
calculations \cite{shif} in the first place. If we are not prepared to
do so we are forced to understand the origin of the divergences that
plague most of the attempts to define even free propagators in radial
gauges and see whether they can be dealt with in a satisfying manner.

In Section~\ref{sec:radgaugecond} we will make the first and decisive
step in this direction by exploring the completeness of the gauge
condition (\ref{eq:fsgaugecond}) and its relation to the field
strength formula and developing a new representation of the gauge
potentials via link operators.

In Section~\ref{sec:radprop} we use this information to relate the
divergences encountered in some of the attempts to define radial
propagators to the renormalization properties of link operators. We
find that even free propagators in radial gauge may feel remnants of
the renormalization properties of closed, gauge invariant Wilson loops.
Surprising as this seems to be superficially it is not impossible
however if we recall that the inhomogeneous term in the gauge
transformation has an explicit $1/g$ factor in it. As a result we are
able to define a regularized radial propagator using the field strength 
formula and established regularization procedures for link operators.

Section~\ref{sec:radcalc} will be devoted to demonstrate the
consistency of our approach by calculating a closed Wilson loop using
our propagator and relating the steps to the equivalent calculation in
Feynman gauge.
 
In Section~\ref{sec:radren} we obtain an explicitly finite version of
our propagator by completing the renormalization program developed for
link operators before we summarize and compare our results to other
approaches in the literature in~\ref{sec:radsum} and shortly discuss
the next steps in the program of establishing a new perturbative
framework in radial gauges which -- although the steps to be performed
are quite straightforward -- we will postpone for a future
publication.

In the following we work in a $D$-dimensional Euclidean space. The vector
potentials are given by
\begin{equation}
A_\mu(x) \equiv A^a_\mu(x) \,t_a
\end{equation}
where $t_a$ denotes the generators of an $SU(N)$ group in the fundamental
representation obeying
\begin{equation}
[t_a,t_b] = i f_{abc} \,t^c
\end{equation}
and
\begin{equation}
\mbox{tr}(t_a t_b) = {1\over 2} \,\delta_{ab}  \,.
\end{equation}
In general the radial gauge condition with respect to $z$ reads
\begin{equation}
(x-z)_\mu A^\mu(x) =0  \,. \label{eq:arbz}
\end{equation} 
For simplicity we take $z=0$ after Section~\ref{sec:radgaugecond}. The
results nevertheless can be easily generalized to arbitrary values of $z$.

%%%%%%%%%%%%%%% gaugecond %%%%%%%%%%%%%%%%%%%%%%%%%%%%%%%%%%%%%%%%%%%%%%%%%
\section{The gauge condition revisited} \label{sec:radgaugecond}

Before we can go ahead and tackle the problem of divergences in the
radial gauge propagator we have to establish a clearer picture of the
uniqueness of the gauge condition we are about to implement. After
all, if we do not succeed to fix the gauge completely we might be
naturally confronted with divergences -- if not at the free level then
later in perturbative calculations. They would be a simple consequence
of the incompleteness of the gauge fixing and the zero modes of the
propagator which would then necessarily be present. This point has
caused a still continuing discussion for the case of axial gauges
(e.g. \cite{leroy}) but is only briefly mentioned in the context of
radial gauges (e.g. \cite{azam}).

Readers who are not interested in the discussion of (in)completeness
of radial gauge conditions might skip the following considerations
without getting lost and start reading again after eq.
(\ref{eq:gaugetrans}).

To clarify the question whether the gauge condition (\ref{eq:arbz}) is
sufficient to completely fix the gauge degrees of freedom we have to
catalogue the gauge transformations $U[B](x)$ which transform an
arbitrary vector potential $B$ into the field $A$ satisfying
(\ref{eq:arbz}).  A gauge condition is complete if $U[B](x)$ is
uniquely determined up to a global gauge transformation. In other
words, we want to find all solutions of
\begin{equation}
  \label{eq:transformations}
  (x-z)_\mu\ U[B](x)\left[B^\mu(x) -{1\over i g}
    \partial^\mu\right]U[B]^{-1}(x) = 0    \,.
\end{equation}
It is easily checked that we have an infinite family of such solutions
which can all be cast in the form of a product of two gauge
transformations of the form
\begin{equation}
  \label{eq:gaugesol}
  U[B](x) = V(z(x)) U[B](z(x),x)  \,.
\end{equation}
Here
\begin{equation}
  \label{eq:gaugetranssol}
  U[B](z(x),x) = {\cal P} \exp i g \int^{z(x)}_x \!d\omega_\mu
  B^\mu(\omega)
\end{equation}
is a link operator whose geometric ingredients are parameterized via
its endpoints $x$ and $z(x)$ and the straight line path $\omega$
between them, ${\cal P}$ denotes path ordering.

\begin{figure}
\begin{center}
  \epsfxsize=\textwidth \epsfbox{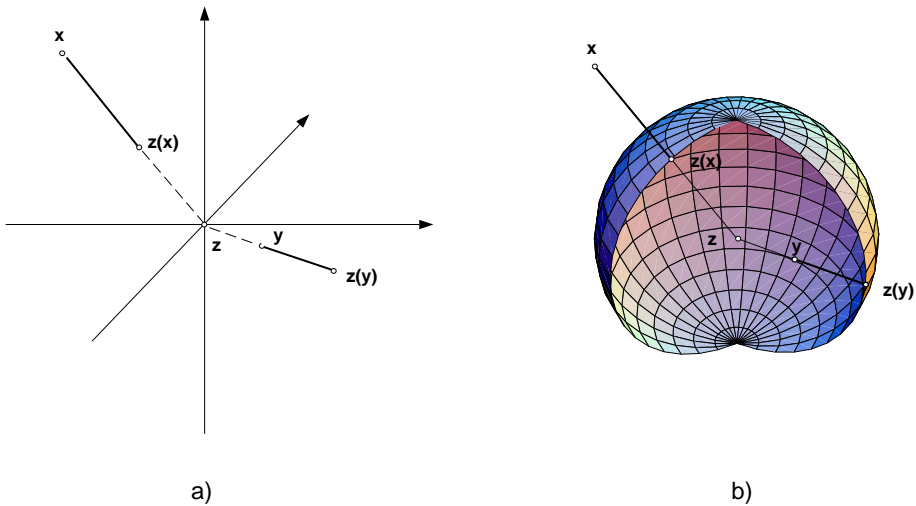}
\caption{a) straight line path in the links for two
  points $x$ and $y$.  b) Example for a spherical hyper-surface $z(x)$.
\label{fig:sphericalgeom}}
\end{center}
\end{figure}

In particular $z(x)$ is the point where a straight line from $z$
through $x$ and a given closed hyper-surface around $z$ intersect.
Since there is a unique relation between these points and the
hyper-surface we will also refer to the hyper-surface itself by $z(x)$.
This geometry is illustrated in Fig.~\ref{fig:sphericalgeom}.
Both the detailed form of $z(x)$ and the local gauge transformation
$V(x)$ are completely unconstrained as long as $(x-z).\partial^x z(x)
= 0$. In short, they parameterize the residual gauge freedom not
eliminated by (\ref{eq:fsgaugecond}). Note that while $V(x)$ is
completely arbitrary the solutions (\ref{eq:gaugesol}) ask only for
its behavior at the given hyper-surface $z(x)$. The simplest and most
intuitive choice for $z(x)$ is a spherical hyper-surface around $z$.
Introducing the appropriate spherical coordinates it becomes obvious
that $V(z(x))$ parameterizes gauge transformations which purely depend
on the angles. Clearly the radial gauge condition (\ref{eq:arbz})
cannot fix the angular dependence of any gauge transformation in
(\ref{eq:transformations}).

To eliminate the residual gauge freedom one has to impose a condition
which is stronger than (\ref{eq:arbz}) and suffices to pin down $V(x)$
up to a global transformation. A possible choice for such a gauge
fixing would be the condition
\begin{eqnarray}
  \label{eq:compgaugefix}
  \Box \left(\int_{z(x)}^x \!d\omega.A(\omega) + \int\!d^4\!y \ {1\over
      \Box}(z(x),y)\  \partial^y.A(y)\right) \equiv 0
\end{eqnarray}
which in addition to the vanishing of the radial component of the
gauge potential also implements a covariant gauge on the hyper-surface
$z(x)$. Such a gauge for arbitrary $z(x)$ would immediately force us
to introduce ghosts into the path integral. Moreover the field
strength formula would also be lost as we will illustrate
below.

There is one exception to these unwanted modifications however, which
may be implemented by contracting the closed surface $z(x)$ to the
point $z$.  Then the influence of $V(x)$ becomes degenerate with a
global transformation and the gauge is completely fixed. Incidentally
this is also the only case which entails the field strength formula.
To see this we use
\begin{eqnarray}
\lefteqn{ \delta U(x,z) = i g \;\Bigg\{ A_\mu(x)U(x,z)dx^\mu 
        - U(x,z)A_\mu(z)dz^\mu + } \nonumber \\
& & - \int^1_0\!\!ds\, [U(x,w_x)F_{\mu\nu}(w_x)U(w_x,z)]
        {dw_x^\mu\over ds} 
        \left ( {dw_x^\nu\over dx^\alpha}dx^\alpha 
                + {dw_x^\nu\over dz^\alpha}dz^\alpha
        \right )
        \Bigg\}    \quad , \nonumber \\
    \label{eq:udiff}
\end{eqnarray}
(see e.g.  \cite{bralic}, \cite{ElGyuVa86}) to differentiate the link
operators in the gauge transformation (\ref{eq:gaugesol}) in order to
find an expression for the radial gauge field:
\begin{eqnarray}
  \label{eq:gaugetrans}
  A^{\mbox{\scriptsize rad}}_\mu(x) & = & U[A](z,x)\left[A_\mu(x) -{1\over i g}
    \partial^x_\mu\right]U[A](x,z) 
\nonumber \\ & = &
   \int\limits^1_0 \!\! ds \, s {d\omega^\nu\over d s}
   \,U[A](z,\omega)\,F_{\nu\mu}(\omega) \,U[A](\omega,z)
\nonumber \\ & = &
   \int\limits^1_0 \!\! ds \, s {d\omega^\nu\over d s}
   \,F_{\nu\mu}^{\mbox{\scriptsize rad}}(\omega)  \,.
\end{eqnarray}
This is nothing but (\ref{eq:fsformula}) for arbitrary $z$ (note that
in this case $\omega = \omega(s)$ is simply given by $\omega(s) = z +
(x-z) s$.)  This simple result is only true since $\partial_\mu z(x)
\equiv \partial_\mu z = 0$. For general $z(x)$ there would be an
additional term in the above formula reflecting the residual gauge
freedom encoded in $V(z(x))$.

%{\bf 
%A simple way to write down this complete radial gauge condition is the
%following expression in terms of the link operator:
%\begin{eqnarray}
%U[A](x,z) :=   {\cal P}\exp\left[
%i g\int\limits^1_0 \!\!ds\, (x-z)_\mu A^\mu(z+s(x-z))
%\right]  \stackrel{!}{=} 1  \quad\mbox{for all x.} 
%\label{eq:unit}\end{eqnarray}
%}
This sets the stage for a further exploration of the radial gauge in a
context where we can be sure of having completely fixed the gauge in
such a way that the field strength formula is guaranteed to be valid.
Before we go on to studying the consequences the above has for the
implementation of propagators we will introduce yet another
representation of the gauge field in this particular complete radial
gauge -- this time solely in terms of link operators.

From now on we will assume the reference point $z$ to be the origin, but
it will always be straightforward to recover the general case without
any ambiguities. We will also suppress the explicit functional
dependence of link operators on the gauge potential $A$ for brevity.

Let us start with a link operator along a straight line path
\begin{equation}
U(x,x') = {\cal P}\exp\left[
i g\int\limits^1_0 \!\!d\omega_\mu A^\mu(\omega)
\right] 
\end{equation}
where now $ \omega(s):= x'+s(x-x')\, $.  According to
(\ref{eq:udiff}) we have
\begin{equation}
  \partial_\mu^x \, U(x,x') = i g \left[ A_\mu(x) - \!\!\int\limits^1_0
    \!\! ds \, s \, {d\omega^\nu\over d s}\ U(x,\omega)\,F_{\nu\mu}(\omega)
    \,U(\omega,x) \right] U(x,x')
\label{eq:difstr}
\end{equation}
which can be used to express the vector potential in terms of the link
operator
\begin{equation}
\lim_{x'\to x} \partial_\mu^x \, U(x,x') = i g \, A_\mu(x) \,.
\end{equation}
In the case at hand the fact that $U(0,x) = 1$ in
any of the $x.A(x) = 0$ gauges allows us to introduce
a new gauge covariant representation
\begin{equation}
A^{\mbox{\scriptsize rad}}_\mu(x) = 
{1\over i g} \lim_{x'\to x} \partial_\mu^x 
\left[ U(0,x)\,U(x,x')\,U(x',0) \right]   \label{eq:gaucov}
\end{equation}
for the Fock-Schwinger gauge field.  It is easy to see that this is
indeed equivalent to the field strength formula as given in
(\ref{eq:fsformula}) and consequently satisfies the same complete
gauge fixing condition (i.e. (\ref{eq:compgaugefix}) for $z(x) \to
z$):
\begin{eqnarray}
A^{\mbox{\scriptsize rad}}_\mu(x) 
&=& 
{1\over i g} \lim_{x'\to x} \partial_\mu^x 
\left[ U(0,x)\,U(x,x')\,U(x',0) \right]
\nonumber\\ &=& 
{1\over i g} \lim_{x'\to x}  
\left[ \partial_\mu^x \,U(0,x) \,U(x,x') \,U(x',0) \right.
\nonumber \\ && \hskip 2cm \left.
+ U(0,x) \,\partial_\mu^x \,U(x,x') \,U(x',0)\right] 
\nonumber\\ &=& 
{1\over i g} \,\partial_\mu^x \,U(0,x) \,U(x,0)
+  U(0,x) \,A_\mu(x)\,U(x,0)
\nonumber\\ &=& 
\int\limits^1_0 \!\! ds \, s x^\nu F^{\mbox{\scriptsize rad}}_{\nu\mu}(sx)  
\label{eq:fistr}
\end{eqnarray}
where the last step uses (\ref{eq:difstr}), mirroring the relations in
(\ref{eq:gaugetrans}) for $z=0$.
 
%%%%%%%%%%%%% Prop %%%%%%%%%%%%%%%%%%%%%%%%%%%%%%%%%%%%%%%%%%%%%%%%%%%%%
\section{The Radial Gauge Propagator} \label{sec:radprop}

Having established the complete gauge fixing we are interested in, it
is now straightforward to devise expressions for the propagator as a
two-point function. According the above we know that
\begin{eqnarray}
  \label{eq:propdef}
  \lefteqn{ \langle A_\mu(x) \otimes A_\nu(y)\rangle_{\mbox{\scriptsize rad}} }
  \nonumber \\& = & \lim_{x'\to x \atop y'\to y}\partial_\mu^x
  \partial_\nu^y \, \langle U(0,x)\,U(x,x')\,U(x',0)\otimes
  U(0,y)\,U(y,y')\,U(y',0) \rangle \nonumber \\ & = & \int\limits^1_0
  \!\! ds \int\limits^1_0 \!\! dt\, sx^\alpha \, ty^\beta \, \langle
  U(0,sx)\,F_{\alpha\mu}(sx) \,U(sx,0)\otimes
  U(0,ty)\,F_{\beta\nu}(ty) \,U(ty,0) \rangle \ \ .
\nonumber \\ &&
\end{eqnarray}

Since we are in a fixed gauge it makes sense to perform a multiplet
decomposition and for instance extract the singlet part of this
propagator. The latter reduces to the free propagator in the limit $g
\to 0$.

\begin{figure}[ht]
  \begin{center}
    \begin{minipage}{5cm}
      \epsfxsize=5cm \epsfysize=5cm \epsfbox{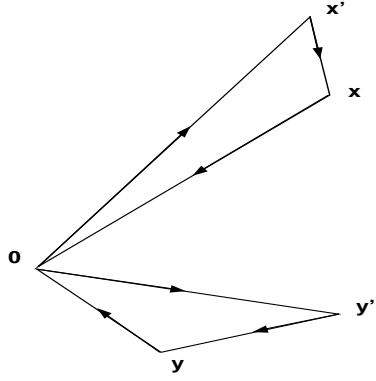}
\end{minipage}
\begin{minipage}{10cm}
\caption{The Wilson loop $W_1$ which is intimately connected with the 
  radial gauge propagator according to ((\ref{propa}), (\ref{w1def}.))}
\label{fig:faech}
\end{minipage}
  \end{center}
\end{figure}

We define
\begin{equation}
  \mbox{tr}\,\langle A_\mu(x) A_\nu(y) \rangle
   =\mbox{tr}(t_a t_b)
  \underbrace{\langle A^a_\mu(x) A^b_\nu(y) \rangle^{\mbox{\scriptsize
        singlet}}}_{{\textstyle
      =:\delta^{ab} D_{\mu\nu}(x,y)}} = {N^2 -1\over 2} \,
  D_{\mu\nu}(x,y)
\end{equation}
to extract
\begin{eqnarray}
  \lefteqn{\langle A^a_\mu(x) A^b_\nu(y) \rangle^{\mbox{\scriptsize
        singlet}}_{\mbox{\scriptsize rad}}
= \delta^{ab} {2\over N^2 -1}
    \mbox{tr}\, \langle A_\mu(x) A_\nu(y) \rangle_{\mbox{\scriptsize
        rad}} = \delta^{ab} {2\over N^2 -1} }
\label{propa}  \\ 
&& \times {1\over (ig)^2} \lim_{x'\to x \atop
  y'\to y}\partial_\mu^x \partial_\nu^y \, \mbox{tr}\,\langle
U(0,x)\,U(x,x')\,U(x',0)\,U(0,y)\,U(y,y')\,U(y',0) \rangle
\,.\nonumber
\end{eqnarray}
Obviously
\begin{eqnarray}
  W_1(x,x',y,y') := {1\over N} \, \mbox{tr}\,\langle
  U(0,x)\,U(x,x')\,U(x',0)\,U(0,y)\,U(y,y')\,U(y',0) \rangle
\label{w1def}\end{eqnarray}
is a gauge invariant Wilson loop.  Its geometry is illustrated in
Fig.~\ref{fig:faech}.

On the other hand, using the second expression in (\ref{eq:propdef})
we have an equivalent representation for the singlet part of radial
gauge propagator via the field strength formula:
\begin{eqnarray}
  \lefteqn{\langle A^a_\mu(x) A^b_\nu(y) \rangle_{\mbox{\scriptsize
        rad}}^{\mbox{\scriptsize singlet}} 
= \delta^{ab} {2\over N^2 -1} }
\label{profi}\\ && \times
\int\limits^1_0 \!\! ds \int\limits^1_0 \!\! dt\, sx^\alpha \,
ty^\beta \, \mbox{tr}\,\langle U(0,sx)\,F_{\alpha\mu}(sx) \,U(sx,0)
\,U(0,ty)\,F_{\beta\nu}(ty) \,U(ty,0) \rangle \,. \nonumber
\end{eqnarray}
Modanese \cite{moda} tried to calculate the free radial gauge
propagator from (\ref{profi}) in a $D$ dimensional
space-time.\footnote{In fact he discussed the Abelian case but this
  makes no difference for free fields.} Unfortunately one gets a
result which diverges in the limit $D\to 4$.

Since we have performed a complete gauge fixing (at least on the
classical level), this comes as a surprise since we certainly do not
expect zero mode problems to come into the game as a possible
explanation and consequently a way out. Does this mean we are trapped
at a dead end or is there another explanation for this seemingly
devastating discovery?

Before we try to answer this question let us briefly recapitulate how
this divergence makes its appearance: Since the right hand side of
(\ref{profi}) is gauge invariant we can choose an arbitrary gauge to
calculate it. For simplicity we take the Feynman gauge with its free
propagator
\begin{equation}
\langle A_\mu^a(x) A_\nu^b(y) \rangle_{\mbox{\scriptsize Feyn}} 
= \delta^{ab}\, D_{\mu\nu}^{\mbox{\scriptsize Feyn}}(x,y)
= -{\Gamma(D/2-1) \over 4\pi^{D/2}}
\,g_{\mu\nu} \,\delta^{ab}\, [(x-y)^2]^{1-D/2}  \,.\label{feynman}
\end{equation}
Using the free field relations $ U(a,b) = 1 $ and $ F_{\mu\nu}=
\partial_\mu A_\nu - \partial_\nu A_\mu $ we get (for more details see
Appendix~\ref{appprop})
\begin{eqnarray}
  \lefteqn{\langle A^a_\mu(x) A^b_\nu(y) \rangle^0_{\mbox{\scriptsize
        rad}}=} \nonumber\\&& = -{\Gamma(D/2-1) \over 4\pi^{D/2}}
  \,\delta^{ab} \int\limits^1_0 \!\! ds \int\limits^1_0 \!\! dt\,
  sx^\alpha \, ty^\beta \nonumber\\&& \phantom{=}\times \left(
    g_{\mu\nu} \partial_\alpha^{sx}\partial_\beta^{ty} +
    g_{\alpha\beta}\partial_\mu^{sx}\partial_\nu^{ty} -
    g_{\alpha\nu}\partial_\mu^{sx}\partial_\beta^{ty} -
    g_{\mu\beta}\partial_\alpha^{sx}\partial_\nu^{ty} \right)
  [(sx-ty)^2]^{1-D/2} \nonumber\\ && = -{\Gamma(D/2-1) \over
    4\pi^{D/2}} \,\delta^{ab} \, \bigg( g_{\mu\nu} [(x-y)^2]^{1-D/2}
  \nonumber\\ && \phantom{mmm} -\partial_\mu^x \int\limits^1_0 \!\! ds
  \, x_\nu\,[(sx-y)^2]^{1-D/2} -\partial_\nu^y \int\limits^1_0 \!\! dt
  \, y_\mu \,[(x-ty)^2]^{1-D/2} \nonumber\\ && \phantom{mmm}
  +\partial_\mu^x \partial_\nu^y \underbrace{\int\limits^1_0 \!\! ds
    \int\limits^1_0 \!\! dt\, x\cdot y \,
    [(sx-ty)^2]^{1-D/2}}_{\textstyle \sim {\textstyle 1\over
      \textstyle 4-D}} \bigg) \,.
\label{divprop}\end{eqnarray}
Thus the radial gauge propagator is singular for arbitrary arguments $x$ and
$y$ --- with one remarkable exception: It is easy to see that it vanishes for
$x=0$ or $y=0$. This is simply a consequence of our task to preserve the
field strength formula (\ref{eq:fsformula}) which forces the vector field
to vanish at the origin (in general at the reference point $z$).

The observation that the radial gauge propagator as calculated
here diverges in four-dimensional space raises the question, whether
it is perhaps impossible to formulate a quantum theory in radial
gauge. This would suggest that the radial gauge condition -- in the
form that facilitates the field strength formula -- is inherently
inconsistent (``unphysical'') in contrast to the general belief that it is 
``very physical'' since it allows to express gauge variant quantities like
the vector potential in terms of gauge invariant ones.  To answer this
question we have to understand where this divergence comes from.  In
the following we will see that for this purpose the complicated
looking Wilson loop representation (\ref{propa}) is much more useful
than the field strength formula (\ref{profi}). (Note, however, that
the result for the free propagator (\ref{divprop}) of course will be
the same.)

It is well-known that Wilson loops need renormalization to make them
well-defined (see e.g.~\cite{korrad} and references therein). The
expansion of an arbitrary Wilson loop
\begin{equation}
W(C) = {1\over N} \, \mbox{tr}\,\left\langle {\cal P}\exp\left[ 
ig\oint_C dx^\mu A_\mu(x) \right] \right\rangle 
\end{equation}
in powers of the coupling constant is given by
\begin{eqnarray}
\lefteqn{
W(C) = 1 + {1\over N} \sum\limits_{n=2}^\infty (ig)^n 
\oint_C \! dx_1^{\mu_1} \ldots \oint_C \! dx_n^{\mu_n}}
\nonumber \\ && \times
\Theta_C(x_1 > \cdots > x_n) \,\mbox{tr}\, 
G_{\mu_1 \ldots \mu_n}(x_1,\ldots,x_n) \label{ptexp}
\end{eqnarray}
where $\Theta_C(x_1 > \cdots > x_n)$ orders the points
$x_1,\ldots,x_n$ along the contour $C$ and
\begin{equation}
G_{\mu_1 \ldots \mu_n}(x_1,\ldots,x_n) :=
\left\langle A_{\mu_1}(x_1) \cdots A_{\mu_n}(x_n) \right\rangle
\end{equation}
are the Green functions.

In general Wilson loops show ultraviolet singularities in any order of
the coupling constant. If the contour $C$ is smooth
(i.e.~differentiable) and simple (i.e.~without self-intersections) the
conventional charge and wave-function renormalization --- denoted by
${\cal R}$ in the following --- is sufficient to make $W(C)$ finite.
We refer to \cite{regul} for more details about renormalization of
regular (smooth and simple) loops.

In our example we must apply the renormalization operation ${\cal R}$
to $W_1$ as given in (\ref{w1def}). This yields
\begin{equation}
  \tilde W_1(x,x',y,y';g_R,\mu,D) = {\cal R} W_1(x,x',y,y';g,D)
  \label{roper}
\end{equation} 
where $W_1(x,x',y,y';g,D)$ is a regularized expression calculated in
$D$ dimensions and $\mu$ is a subtraction point introduced by the
renormalization procedure ${\cal R}$. For the purpose of the present
work the only important relation is
\begin{equation}
g_R = \mu^{(D-4)/2} g + o(g^3)  \,. \label{grug}
\end{equation}

While the operation ${\cal R}$ is sufficient to make regular loops
well-defined, new divergences appear if the contour $C$ has cusps or
self-intersections. The renormalization properties of such loops are
discussed in \cite{brandt} and \cite{brandt2}. While the singularities
of regular loops appear at the two-loop level (order $g^4$ in
(\ref{ptexp})) cusps and cross points cause divergences even in
leading (non-trivial) order $g^2$.

%One can give an intuitive picture for this singularities by observing
%that a Wilson loop along the contour $C$ can be described as a
%one-dimensional fermion field living on $C$. If the contour has an
%cusp ????

Since $W_1$ is indeed plagued by cusps and self-intersections a second
renormalization operation must be carried out to get a renormalized
expression $W_1^R$ from the bare one $W_1$: According to \cite{brandt}
each cusp is multiplicatively renormalizable with a renormalization
factor $Z$ depending on the cusp angle. In our case we have four cusps
with angles
\begin{eqnarray}
\alpha  &:=& \angle (x-x',-x) \,, \\
\alpha' &:=& \angle (x',x-x') \,, \\
\beta   &:=& \angle (y-y',-y) \,, \\
\beta'  &:=& \angle (y',y-y') \,.
\end{eqnarray}
The cross point at the origin introduces a mixing between $W_1$ and
\begin{eqnarray}
\lefteqn{W_2(x,x',y,y') :=}\nonumber\\
&&\left\langle
{1\over N} \,\mbox{tr}\left[ U(0,x)\,U(x,x')\,U(x',0)\right] \,
{1\over N} \,\mbox{tr}\left[ U(0,y)\,U(y,y')\,U(y',0)\right] 
\right\rangle \,.
\end{eqnarray}
Again the divergences appearing here are functions of the angles
\begin{equation}
\left.
\begin{array}{lcl}
\gamma_{xx'} &:=& \angle (-x,x')  \\
\gamma_{yy'} &:=& \angle (-y,y')  \\
\gamma_{xy} &:=& \angle (-x,-y)  \\
\gamma_{x'y'} &:=& \angle (x',y')  \\
\gamma_{x'y} &:=& \angle (x',-y)  \\
\gamma_{xy'} &:=& \angle (-x,y')
\end{array}
\right\} \vec\gamma \,.
\end{equation} 
The renormalized Wilson loop $W_1^R$ is given by
\begin{eqnarray}
\lefteqn{W_1^R(x,x',y,y';g_R,\mu,
\bar C_\alpha,\bar C_{\alpha'},\bar C_\beta,\bar C_{\beta'},
\bar C_{\vec\gamma})}
\nonumber\\ &&
= \lim_{D\to 4} 
\, Z(\bar C_\alpha,g_R,\mu;D) \, Z(\bar C_{\alpha'},g_R,\mu;D) 
\, Z(\bar C_\beta,g_R,\mu;D) \, Z(\bar C_{\beta'},g_R,\mu;D)
\nonumber\\ && \phantom{=\lim}
\times \left[ Z_{11}(\bar C_{\vec\gamma},g_R,\mu;D) \,
\tilde W_1(x,x',y,y';g_R,\mu,D) 
\right. \nonumber\\ && \phantom{=\lim \times [} \left.
{}+Z_{12}(\bar C_{\vec\gamma},g_R,\mu;D) \,
\tilde W_2(x,x',y,y';g_R,\mu,D)
\right] 
\nonumber\\ &&
=: \lim_{D\to 4} \bar W_1(x,x',y,y';g_R,\mu,
\bar C_\alpha,\bar C_{\alpha'},\bar C_\beta,\bar C_{\beta'},
\bar C_{\vec\gamma};D) \label{barw}
\end{eqnarray}
where the second renormalization procedure introduces new
subtraction points $\bar C_\sigma$ (c.f.~\cite{korrad} and
\cite{brandt} for more details). Of course different renormalization
procedures are possible and so the $Z$ factors are not unique. We will
return to this point in Section~\ref{sec:radren} where we specify a
renormalization operation which is appropriate for our purposes.

The observation that Wilson loops with cusps and/or cross points show
additional divergences has an important consequence for our radial
gauge propagator as given in (\ref{propa}): Even the free propagator
needs renormalization! This provides a natural explanation for the
fact that a naive calculation of this object yields an ultraviolet
divergent result \cite{moda}.  Note that the usual divergences of
Wilson loops which are removed by ${\cal R}$, like e.g.~vertex
divergences, appear at $o(g^4)$ and thus do not contribute to the free
part of the radial gauge propagator, while the cusp singularities indeed 
contribute since they appear at $o(g^2)$ and affect the free field case due to
the factor $1/g^2$ in (\ref{propa}).

Now we are able to answer the question whether the radial gauge is
``unphysical'' or ``very physical''. It is just its intimate relation
to physical, i.e.~gauge invariant, quantities which makes the gauge
propagator --- even the free one --- divergent. One might cast the
answer in the following form: {\it The propagator diverges because of
  --- and not contrary to --- the fact that the radial gauge is ``very
  physical''}.

Consequently the next questions are:
\begin{itemize}
\item[---] Is there any use for a divergent expression for the free
  propagator?  Especially: Can we use it to perform (dimensionally
  regularized) loop calculations?
\item[---] Can one find a renormalization program which yields a
  finite radial gauge propagator?
\end{itemize}

In the next Section we will perform a one loop calculation of a Wilson
loop using the radial gauge propagator (\ref{divprop}) and compare the
dimensionally regularized result with a calculation in Feynman gauge.

In Section~\ref{sec:radren} we will explicitly demonstrate that the
renormalization program for link operators carries over and allows to
derive a finite result for the radial propagator and contrast its
properties and use to the regularized version.

%%%%%%%%%%%%%%%%%%%%%% calc %%%%%%%%%%%%%%%%%%%%%%%%%%%%%%%%%%%%%%
\section{Calculating a Wilson Loop in Radial Gauge} \label{sec:radcalc}

\begin{figure}[ht]
\begin{center}
\begin{minipage}{5cm}
\epsfysize=5cm
\epsfbox{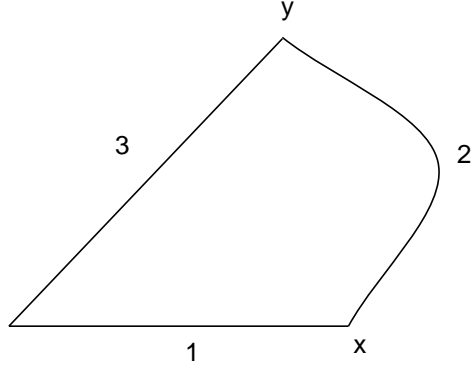}
\end{minipage}
\begin{minipage}{10cm}
\caption{A Wilson loop with two straight line parts.} \label{fig:drop}
\end{minipage}
\end{center}
\end{figure}

We choose the path
\begin{equation}
\ell : z(\sigma) = \left\{
\begin{array}{lclcl}
\sigma x &,& \sigma\in [0,1] &,& x\in{\mathbb R}^D  \\
w(\sigma -1) &,& \sigma\in [1,2] &,& w(0)=x,\, w(1)=y  \\
(3-\sigma)\,y &,& \sigma\in [2,3] &,& y\in{\mathbb R}^D
\end{array}
\right. \label{looppath}
\end{equation}
It is shown in Fig.~\ref{fig:drop}. The line $w(\sigma-1)$ is supposed
to be an arbitrary curve connecting $x$ and $y$.

First we will perform the calculation of this Wilson loop in Feynman gauge. 
Using (\ref{feynman}) we get in leading order of the coupling constant
\begin{eqnarray}
W(\ell) &=&
{1\over N} \, \mbox{tr}\,\left\langle {\cal P}\exp\left[ 
ig\oint_\ell dz^\mu A_\mu(z) \right] \right\rangle 
\nonumber \\ &\approx&
1 + (ig)^2 {N^2 -1\over 2N} 
\int\limits^3_0 \!\! d\sigma \!\int\limits^3_0 \!\! d\tau \, 
\Theta(\sigma-\tau)\,\dot z^\mu(\sigma) \, \dot z^\nu(\tau) \, 
D_{\mu\nu}^{\mbox{\scriptsize Feyn}}(z(\sigma),z(\tau))  
\nonumber \\ &=&
1 + (ig)^2 {N^2 -1\over 2N} {1\over 2}
\underbrace{\int\limits^3_0 \!\! d\sigma \!\int\limits^3_0 \!\! d\tau \, 
\dot z^\mu(\sigma) \, \dot z^\nu(\tau) \, 
D_{\mu\nu}^{\mbox{\scriptsize Feyn}}(z(\sigma),z(\tau))} 
_{\textstyle =: I_f}  \,. \label{feynres}
\end{eqnarray}
To get rid of the $\Theta$-function we have exploited the symmetry
property of two-point Green functions
\begin{equation}
D_{\mu\nu}^{\mbox{\scriptsize Feyn}}(x,y)
= D_{\nu\mu}^{\mbox{\scriptsize Feyn}}(y,x)  \,.
\end{equation}
Decomposing the contour $\ell$ according to (\ref{looppath}) we find
that the Feynman propagator in (\ref{feynres})) connects each segment
of $\ell$ with itself and with all the other segments.  Thus $I_f$ is
given by
\begin{equation}
I_f = \sum\limits_{A = 1}^3 \sum\limits_{B = 1}^3\,(A,B)
\end{equation}
where $(A,B)$ denotes the contribution with propagators connecting
loop segments $A$ and $B$ (c.f.~Fig.~\ref{fig:drop}), e.g.
\begin{eqnarray}
(1,2) &=& 
\int\limits^1_0 \!\! d\sigma \!\int\limits^1_0 \!\! d\tau \,
x^\mu \, \dot w^\nu(\tau) \, 
D_{\mu\nu}^{\mbox{\scriptsize Feyn}}(\sigma x,w(\tau))
\nonumber\\ &=&
-{\Gamma(D/2-1) \over 4\pi^{D/2}}
\int\limits^1_0 \!\! d\sigma \!\int\limits^1_0 \!\! d\tau \,
x^\mu \, \dot w_\mu(\tau) \, [(\sigma x-w(\tau))^2]^{1-D/2}  \,.
\end{eqnarray}

Next we will evaluate the same Wilson loop in radial gauge. Clearly
the first and the third part of the path do not contribute if the
radial gauge condition $x_\mu A^\mu(x) = 0$ holds. We insert the free
propagator
\begin{eqnarray}
\langle A^a_\mu(x) A^b_\nu(y) \rangle^0_{\mbox{\scriptsize rad}}
=: \delta^{ab} D^0_{\mu\nu}(x,y)  
\end{eqnarray}
from (\ref{divprop}) into
\begin{eqnarray}
W(\ell) &=&
{1\over N}\, \mbox{tr}\,\left\langle {\cal P}\exp\left[ 
ig\int\limits_0^1 \!\! d\sigma \,\dot w_\mu(\sigma) A^\mu(w(\sigma))
\right]\right\rangle
\nonumber\\ &\approx&
1 + (i g)^2  {N^2 -1\over 2N}
%\nonumber\\ && \times 
{1\over 2}
\underbrace{\int\limits^1_0 \!\! d\sigma \!
\int\limits^1_0 \!\! d\tau \, 
\dot w^\mu(\sigma)\, \dot w^\nu(\tau) \, 
D_{\mu\nu}^0(w(\sigma),w(\tau))} 
_{\textstyle =: I_r} 
\label{wilir}\end{eqnarray}
and observe that 
\begin{equation}
\dot w_\mu(\sigma) \,\partial^\mu_{w(\sigma)} = {d\over d\sigma} \,.
\end{equation}
Thus the integral in (\ref{wilir}) reduces to
\begin{eqnarray}
\lefteqn{I_r = 
-{\Gamma(D/2-1) \over 4\pi^{D/2}} 
\Bigg[
   \int\limits^1_0 \!\! d\sigma \int\limits^1_0 \!\! d\tau \, 
   \dot w_\mu(\sigma)\, \dot w^\mu(\tau) \, 
[(w(\sigma)-w(\tau))^2]^{1-D/2}}
\nonumber\\&& \phantom{-{\Gamma(D/2-1) \over 4\pi^{D/2}}}
 {}+ \int\limits^1_0 \!\! ds \int\limits^1_0 \!\! dt\, 
   \Big(
        w_\mu(1)\,w^\mu(1)\,[(sw(1)-tw(1))^2]^{1-D/2}
\nonumber\\&& 
\phantom{-MM(D/2-1)\int\limits^1_0\!\! ds\int\limits^1_0\!\!dt\,}
        {}+ w_\mu(0)\,w^\mu(0)\,[(sw(0)-tw(0))^2]^{1-D/2}
\nonumber\\&& 
\phantom{-MM(D/2-1)\int\limits^1_0\!\! ds\int\limits^1_0\!\!dt\,}
        {}-w_\mu(1)\,w^\mu(0)\,[(sw(1)-tw(0))^2]^{1-D/2}
\nonumber\\&& 
\phantom{-MM(D/2-1)\int\limits^1_0\!\! ds\int\limits^1_0\!\!dt\,}
        {}-w_\mu(0)\,w^\mu(1)\,[(sw(0)-tw(1))^2]^{1-D/2}
   \Big)
\nonumber\\&& \phantom{-{\Gamma(D/2-1) \over 4\pi^{D/2}}}
 {}- \int\limits^1_0 \!\! ds \int\limits^1_0 \!\! d\tau \,
   \dot w_\mu(\tau)
   \Big(
        w^\mu(1) [(sw(1)-w(\tau))^2]^{1-D/2}
\nonumber\\&& 
\phantom{-MMMM(D/2-1)\int\limits^1_0\!\! ds\int\limits^1_0\!\!dt\,}
        {}- w^\mu(0) [(sw(0)-w(\tau))^2]^{1-D/2}
   \Big)
\nonumber\\&& \phantom{-{\Gamma(D/2-1) \over 4\pi^{D/2}}}
 {}- \int\limits^1_0 \!\! dt \int\limits^1_0 \!\! d\sigma \,
   \dot w_\mu(\sigma)
   \Big(
        w^\mu(1) [(w(\sigma)-tw(1))^2]^{1-D/2}
\nonumber\\&& 
\phantom{-MMMM(D/2-1)\int\limits^1_0\!\! ds\int\limits^1_0\!\!dt\,}     
        {}- w^\mu(0) [(w(\sigma)-tw(0))^2]^{1-D/2}
   \Big)
\Bigg]
\label{lengthy}\\&&
= (2,2) + (3,3) + (1,1) + (3,1) + (1,3) + (3,2) + (1,2) + (2,3) +
(2,1) \,.  \nonumber
\end{eqnarray} 
A careful analysis of (\ref{lengthy}) shows that it exactly coincides
with the Feynman gauge calculation. This is expressed in the last line
where we have denoted which parts of the loop are connected by the
Feynman gauge propagator to reproduce (\ref{lengthy}) term by term.
Thus using the radial gauge propagator as given in (\ref{divprop})
yields the same result as the calculation in Feynman gauge. Finally
this regularized expression has to be renormalized. This can be
performed without any problems according to \cite{brandt}. Since we
are not interested in the Wilson loop itself but in the comparison of
the results obtained in radial and Feynman gauge, we will not
calculate the renormalized expression for $W(\ell)$. 

However a qualitative discussion of the renormalization properties of
$W(\ell)$ is illuminating.  By construction $W(\ell)$ has at least a
cusp at the origin. (Other cusps are possible at $x$ or $y$ or along
the line parameterized by $w$, but are not important for our
considerations.) To give the right behavior of the Wilson loop the
calculation of $W(\ell)$ in an arbitrary gauge must reproduce the cusp
singularity. Usually the parameter integrals in the vicinity of the
cusp do the job. For gauge choices where the propagator do not vanish
in the vicinity of the origin this is automatically achieved. Let us
assume for a moment that it is possible to construct a {\it finite\/}
radial gauge propagator obeying the field strength formula
(\ref{eq:fsformula}) and therefore have trivial gauge factors along
radial lines.
%, i.e.~the complete gauge condition (\ref{eq:unit}).
Of course this is nothing but saying that there are no contributions
form parts 1 and 3 of the loop, i.e.~in the vicinity of the origin.
Since the propagator is assumed to be finite, there are no singular
integrals corresponding to the cusp at the origin.  Thus a finite
radial gauge propagator cannot reproduce the correct behavior of the
Wilson loop. In turn we conclude that {\it a singular radial gauge
  propagator is mandatory\/} to get the right renormalization properties
of Wilson loops.

However, as we will demonstrate in the next Section, the
renormalization procedure for Wilson loops can be used to devise a
consistent renormalization program for the radial gauges considered
here. We will apply it to write down a finite version of the free
radial propagator.  The generalization to higher orders in
perturbation theory is straightforward. According to our
considerations given above we shall show that the renormalized, thus
finite version of the free radial propagator is not suitable as an
input to perturbative calculations.

%%%%%%%%%%%%%%%%% ren %%%%%%%%%%%%%%%%%%%%%%%%%%%%%%%%%%%%%%%%%%%%%%%%%%%%
\section{The Renormalized Free Propagator} \label{sec:radren}

We define the renormalized radial gauge propagator by
\begin{equation}
\langle A^a_\mu(x) A^b_\nu(y) 
\rangle_R^{\mbox{\scriptsize singlet}} := 
\lim_{D\to 4} \delta^{ab} {2N\over N^2 -1} 
{1\over (ig_R)^2} \mu^{D-4}
\lim_{x'\to x \atop y'\to y}\partial_\mu^x \partial_\nu^y \,
\bar W_1(x,x',y,y';D) \label{propren}
\end{equation}
where we have suppressed most of the other variables on which $\bar
W_1$ depends (see (\ref{barw})).

From now on we will concentrate on the calculation of the renormalized
free propagator $\langle A^a_\mu(x) A^b_\nu(y) \rangle_R^0$. The
details of the renormalization program are presented in
Appendix~\ref{app:renprog}.  Of course the procedure is closely
connected to the renormalization of cusp singularities of Wilson
loops. The result is
\begin{eqnarray}
\lefteqn{\langle A^a_\mu(x) A^b_\nu(y) \rangle_R^0 =} 
\\ &&
=\lim_{D\to 4} \left(
\delta^{ab} \mu^{D-4} \partial_\mu^x \partial_\nu^y 
\left(
{1\over 4\pi^2} {1\over 4-D} (\pi-\gamma_{xy})\cot\gamma_{xy} 
\right)
+\langle A^a_\mu(x) A^b_\nu(y) \rangle^0_{\mbox{\scriptsize rad}}
\right)
\nonumber\end{eqnarray}

Before discussing some properties of the renormalized free propagator
we shall show that the counter term
\begin{equation}
C_{\mu\nu}^{ab}(x,y) :=
\delta^{ab} \mu^{D-4} \partial_\mu^x \partial_\nu^y 
\left(
{1\over 4\pi^2} {1\over 4-D} (\pi-\gamma_{xy})\cot\gamma_{xy} 
\right)
\end{equation}
exactly cancels the divergence of the propagator (\ref{divprop}),
i.e.~that $\langle A^a_\mu(x) A^b_\nu(y) \rangle_R^0$ really is
finite. To this end we use some technical results derived in
Appendix~\ref{siint}. The divergent part of the propagator
(\ref{divprop}) is given by
\begin{equation}
U_{\mu\nu}^{ab}(x,y):=-{\Gamma(D/2-1) \over 4\pi^{D/2}} \,\delta^{ab} \,
\partial_\mu^x \partial_\nu^y 
\int\limits^1_0 \!\! ds \int\limits^1_0 \!\! dt\, x\cdot y \, 
[(sx-ty)^2]^{1-D/2} \,.
\end{equation}
Using (\ref{appi2}) and (\ref{i2neq}) we find
\begin{eqnarray}
U_{\mu\nu}^{ab}(x,y) &=&
{\Gamma(D/2-1) \over 4\pi^{D/2}} \,\delta^{ab} \,
\partial_\mu^x \partial_\nu^y \, I_2(x,-y)
\nonumber\\ &=&
{\Gamma(D/2-1) \over 4\pi^{D/2}} \,\delta^{ab} \,
\partial_\mu^x \partial_\nu^y  
\left({1\over 4-D} \,(\pi-\gamma_{xy}) \cot(\pi-\gamma_{xy}) \; 
+\mbox{finite} \right)
\nonumber\\ &=&
-{1 \over 4\pi^2} \,\delta^{ab} \,
\partial_\mu^x \partial_\nu^y  
\left({1\over 4-D} \,(\pi-\gamma_{xy}) \cot\gamma_{xy} \right) \; 
+\mbox{finite}
\end{eqnarray}
and thus
\begin{equation}
U_{\mu\nu}^{ab}(x,y) + C_{\mu\nu}^{ab}(x,y) = \mbox{finite} \,.
\end{equation}

Note that if one tries to guess a finite expression like $\langle
A^a_\mu(x) A^b_\nu(y) \rangle_R^0$ one would have to introduce a scale
$\mu$ by hand without interpretation. In our derivation this scale
appears naturally as the typical renormalization scale of the ${\cal
  R}$ operation.

The counter term $C_{\mu\nu}^{ab}(x,y)$ has some interesting properties. 
It is symmetric with respect to an exchange of all variables 
and it obeys the gauge condition
\begin{equation}
x^\mu C_{\mu\nu}^{ab}(x,y) = 0 = C_{\mu\nu}^{ab}(x,y) \,y^\nu \,.  
\label{courad}\end{equation}
Thus $\langle A^a_\mu(x) A^b_\nu(y) \rangle_R^0$ is finite in the
limit $D\to 4$ but still can be interpreted as a gluonic two-point
function which fulfills the radial gauge condition
\begin{equation}
x^\mu \,\langle A^a_\mu(x) A^b_\nu(y) \rangle_R^0 =0 \,.
\end{equation}
However the counter term $C_{\mu\nu}^{ab}(x,y)$ and thus also 
$\langle A^a_\mu(x) A^b_\nu(y) \rangle_R^0$ is ill-defined at the origin
%violating the complete gauge fixing condition (\ref{eq:unit}) and the
and hence conflicts with the field strength formula
(\ref{eq:fsformula}). Note that the regularized propagator in contrast
to the renormalized propagator is well defined and vanishes if one of
its arguments approaches zero, as pointed out after
eq.~(\ref{divprop}).  We therefore conclude that we may use the
regularized propagator in perturbative calculations and can be ensured
to preserve relations like the field strength formula or
eqs.~(\ref{eq:fistr}) or (\ref{eq:propdef}) throughout the
calculation. Although the counterterms are not well defined at the
reference point itself -- a property the radial gauge propagator
simply inherits from renormalizing the cusp singularity of the
underlying Wilson line -- physical (gauge invariant) quantities are
not affected, they are rendered finite and unambiguous.

%%%%%%%%%%%%%%%% Summary %%%%%%%%%%%%%%%%%%%%%%%%%%%%%%%%%%%%%%%%%%%%%
\section{Summary and Outlook} \label{sec:radsum}

In this article we have shown how to calculate the radial gauge
propagator in a $D$-dimensional space using Wilson loops. As
discovered in \cite{moda} the free propagator diverges in
four-dimensional space. We were able to explain this singular behavior
by studying the properties of associated Wilson loops. Furthermore we
have shown that the free propagator, in spite of being divergent in
four dimensions, can be used for perturbative calculations in a
(dimensionally) regularized framework and that the result for a
gauge invariant quantity agrees with the calculation in Feynman gauge.
Finally we have presented a renormalization procedure for the radial
gauge propagator and calculated the explicit form of the renormalized
free propagator.  We have pointed out that any version of the radial
propagator which is finite in four-dimensional space at least cannot
reproduce the correct renormalization properties of Wilson loops with
cusps at the reference point $z$.

It is instructive to compare the radial gauge propagators as presented here
with other approaches: As discussed in Section~\ref{sec:radgaugecond} the
radial gauge condition (\ref{eq:fsgaugecond}) does not completely fix the
gauge degrees of freedom. Thus the field strength formula
\begin{equation}
A_\mu(x) = \int\limits^1_0 \!\! ds \, s x^\nu F_{\nu\mu}(sx) \label{fistrag}
\end{equation}
is not the only solution of the system of equations\footnote{For simplicity
we discuss the QED case here. Aiming at an expression for the free gauge 
propagator this is no restriction of generality. For non-Abelian gauge groups
c.f.~\cite{azam}.}
\begin{equation}
\left\{\begin{array}{l}
x_\mu A^\mu(x) =0 \,, \\
F_{\mu\nu}(x) = 
\partial^x_\mu A_\nu(x) - \partial^x_\nu A_\mu(x) \,.
\end{array} \right.
\end{equation}
One might add a function \cite{moda}
\begin{equation}
A_\mu^0(x) = \partial^x_\mu f(x)
\end{equation}
to (\ref{fistrag}) where $f$ is an arbitrary homogeneous function of
degree 0.  However any $A_\mu^0(x)$ added in in order to modify
(\ref{fistrag}) is necessarily singular at the origin. Hence
regularity at the origin may be used as a uniqueness condition
\cite{cron}. If we relax this boundary condition other solutions are
possible, e.g.
\begin{equation}
\bar A_\mu(x) = -\int\limits^\infty_1 \!\! ds \, s x^\nu F_{\nu\mu}(sx) 
\label{fistrinf}\end{equation}
where we must assume that the field strength vanishes at infinity.
While (\ref{fistrag}) is the only solution which is regular at the
origin, (\ref{fistrinf}) is regular at infinity. Ignoring boundary
conditions for the moment one can construct a radial gauge propagator
by \cite{menot}
\begin{equation}
{1\over 2} \left( G_{\mu\nu}(x,y) + G_{\nu\mu}(y,x) \right)
\label{menprop}
\end{equation}
with
\begin{equation}
G_{\mu\nu}(x,y) := 
-\int\limits^1_0 \!\! ds \, s x^\alpha 
\int\limits^\infty_1 \!\! dt \, t y^\beta
\left\langle F_{\alpha\mu}(sx) F_{\beta\nu}(ty) \right\rangle  \,.
\end{equation}
It turns out that this propagator is finite in four dimensions.
However the price one has to pay is that boundary conditions are
ignored and thus the object ``lives'' in the restricted space
${\mathbb R}^4 \setminus\{0\}$ and not in ${\mathbb R}^4$ anymore. In
our approach we insist on the field strength formula (\ref{fistrag})
widely used in operator product expansions \cite{shif} and on the
regular behavior of vector potentials at the origin \cite{cron}.  One
might use the propagator (\ref{menprop}) to calculate the
$g^2$-contribution to the Wilson loop on the contour (\ref{looppath}).
It is easy to check that the result differs from the one obtained in
(\ref{wilir}), (\ref{lengthy}). Clearly this is due to the fact that
(\ref{menprop}) is ill-defined at the origin.

In the above, all calculations were performed in Euclidean space. In
Minkowski space Wilson loops show additional divergences if part of
the contour coincides with the light cone \cite{korkor}. Thus we
expect the appearance of new singularities also for the radial
propagator, at least if one or both of its arguments are light-like.
Further investigation is required to work out the properties of the
radial gauge propagator in Minkowski space.

To formulate perturbation theory in a specific gauge the knowledge of
the correct free propagator is only the first step. In addition one
has to check the decoupling of Faddeev-Popov ghosts in radial gauge
which is suggested by the algebraic nature of the gauge condition.
However the still continuing discussion about temporal and axial
gauges might serve as a warning that the decoupling of ghosts for
algebraic gauge conditions is far from being trivial
(c.f.~\cite{cheng}, \cite{leroy} and references therein). To prove (or
disprove) the decoupling of ghosts in radial gauge we expect that our
Wilson loop representation of the propagator is of great advantage
since it yields the possibility to calculate higher loop contributions
in two distinct ways: On the one hand one might use the Wilson loop
representation to calculate the full radial propagator up to an
arbitrary order in the coupling constant. The appropriate Wilson loop
can be calculated in any gauge, e.g.~in a covariant gauge. On the
other hand the radial propagator might be calculated according to
Feynman rules. Since the results should coincide this might serve as a
check for the validity and completeness of a set of radial gauge
Feynman rules.

{\bf Acknowledgments: }

HW wants to thank Alex Kovner for his invaluable patience in his role
as a testing ground of new ideas. SL thanks Professor Ulrich Heinz for
valuable discussions and support.  During this research SL was
supported in part by Deutsche Forschungsgemeinschaft and
Bundesministerium f\"ur Bildung, Wissenschaft, Forschung und
Technologie.  HW was supported by the U.S. Department of Energy under
grants No. DOE Nuclear DE--FG02--87ER--40328 and by the Alexander von
Humboldt Foundation through their Feodor Lynen program.

\begin{appendix}
%%%%%%%%%%%%%%%%% Appendix 1 %%%%%%%%%%%%%%%%%%%%%%%%%%%%%%%%%%%%%%%%%%
\section{Derivation of the Free Radial Propagator} \label{appprop}

The free radial propagator derived form the field strength formula
shows a divergence in $D=4$, as already indicated in
section~\ref{sec:radprop}, eq. (\ref{divprop}). Here we give the
details of the algebra leading to this conclusion.

The following relations summarize the steps carried out in the
calculation below:
\begin{eqnarray}
&& \hskip 2cm
x_\mu \partial_x^\mu = \vert x\vert \,\partial_{\vert x\vert} \,,
\\
T_{\mu\nu}(x,y) &:=& 
x^\alpha  y^\beta  \left( 
g_{\mu\nu} \partial_\alpha^{x}\partial_\beta^{y}
+ g_{\alpha\beta}\partial_\mu^{x}\partial_\nu^{y}
- g_{\alpha\nu}\partial_\mu^{x}\partial_\beta^{y}
- g_{\mu\beta}\partial_\alpha^{x}\partial_\nu^{y}
\right) 
%\nonumber
\\ &=&
g_{\mu\nu} \partial_{\vert x\vert} \partial_{\vert y\vert}
\vert x\vert \,\vert y\vert
-\partial_\mu^x \,x_\nu \,\partial_{\vert y\vert} \vert y\vert
-\partial_\nu^y \,y_\mu \,\partial_{\vert x\vert} \vert x\vert
+\partial_\mu^x \partial_\nu^y \,x\cdot y \,. \nonumber %\\ &&
\end{eqnarray}
Introducing $\hat x :=x/\vert x\vert$ and $u=s\vert x\vert$, we have
for arbitrary $f$:
\begin{eqnarray}
&& \hskip 2cm  sx_\alpha \,\partial_\beta^{sx} f(sx) = x_\alpha \,\partial_\beta^x
  f(sx)  \,,
\\ &&
  \partial_{\vert x\vert}\int\limits_0^1 \!\! ds \, \vert x\vert f(sx)
  = \partial_{\vert x\vert} \int\limits_0^1 \!\! ds \, \vert x\vert
  f(s\vert x\vert \hat x) = \partial_{\vert x\vert}
  \int\limits_0^{\vert x\vert} \!\! du \, f(u \hat x) = f(\vert x\vert
  \hat x) = f(x) \,. \nonumber \\ &&
\end{eqnarray}

We get
\begin{eqnarray}
\lefteqn{{}-{4\pi^{D/2} \over \Gamma(D/2-1)} \,
\langle A^a_\mu(x) A^b_\nu(y) \rangle^0_{\mbox{\scriptsize rad}}=}
\nonumber\\&&
= \delta^{ab}
\int\limits^1_0 \!\! ds \int\limits^1_0 \!\! dt\,
T_{\mu\nu}(sx,ty) \,[(sx-ty)^2]^{1-D/2}
\nonumber\\ &&
= \delta^{ab} \,T_{\mu\nu}(x,y) 
\int\limits^1_0 \!\! ds \int\limits^1_0 \!\! dt\,
[(sx-ty)^2]^{1-D/2}
\nonumber\\ &&
= \delta^{ab} 
\left(
g_{\mu\nu} \partial_{\vert x\vert} \partial_{\vert y\vert}
\vert x\vert \,\vert y\vert
-\partial_\mu^x \,x_\nu \,\partial_{\vert y\vert} \vert y\vert
-\partial_\nu^y \,y_\mu \,\partial_{\vert x\vert} \vert x\vert
+\partial_\mu^x \partial_\nu^y \,x\cdot y 
\right)
\nonumber\\&& \phantom{=}\times
\int\limits^1_0 \!\! ds \int\limits^1_0 \!\! dt\,
[(sx-ty)^2]^{1-D/2}
\nonumber\\ &&
= \delta^{ab} 
\bigg(
g_{\mu\nu} [(x-y)^2]^{1-D/2}
\nonumber\\ && \phantom{mmm}
-\partial_\mu^x \int\limits^1_0 \!\! ds \,x_\nu \, [(sx-y)^2]^{1-D/2}
-\partial_\nu^y \int\limits^1_0 \!\! dt \,y_\mu \, [(x-ty)^2]^{1-D/2}
\nonumber\\ && \phantom{mmm}
+\partial_\mu^x \partial_\nu^y 
\underbrace{\int\limits^1_0 \!\! ds \int\limits^1_0 \!\! dt \, x\cdot y \, 
[(sx-ty)^2]^{1-D/2}}_{\textstyle \sim {\textstyle 1\over \textstyle 4-D}} 
\bigg) \,.
\end{eqnarray}
The divergent part of the double integral in the last line can be found in 
Appendix~\ref{siint}. At the moment however the exact form of the divergence
is not important. 

%%%%%%%%%%%%%%%%%%%%5 Appendix 2 %%%%%%%%%%%%%%%%%%%%%%%%%%%%%%%%%%%%%%%
\section{Renormalization Program for the Free Propagator}
\label{app:renprog} 

In Section~\ref{sec:radren} we discussed the effect of renormalization
on the free radial propagator. Here derive in detail the appropriate
renormalization procedure starting form the renormalization properties
of Wilson lines.  Only a few of the many possible renormalization
constants will contribute to the final result.

Since in the relation between the propagator and the appropriate
Wilson loop (\ref{propa}) a factor $1/g^2$ is involved all quantities
especially all the $Z$'s and $W$'s of eq.~(\ref{barw}) have to be
calculated up to $o(g_R^2)$. We have
\begin{eqnarray}
  \tilde W_i &=& 1 + (ig_R)^2 \delta\tilde W_i + o(g_R^4) \qquad
  (i=1,2)\,, \\ Z &=& 1 + (ig_R)^2 \delta Z + o(g_R^4) \,,\\ Z_{11}
  &=& 1 + (ig_R)^2 \delta Z_{11} + o(g_R^4) \,,\\ Z_{12} &=& 0 +
  (ig_R)^2 \delta Z_{12} + o(g_R^4)
\end{eqnarray}

yielding
\begin{eqnarray}
\lefteqn{\langle A^a_\mu(x) A^b_\nu(y) \rangle_R^0 = 
\lim_{D\to 4} \delta^{ab} {2N\over N^2 -1} \mu^{D-4}}
\label{freepr} \\ && \times
\lim_{x'\to x \atop y'\to y}\partial_\mu^x \partial_\nu^y \,
\left[ 
  \begin{array}{c}
\delta Z(\bar C_\alpha) +\delta Z(\bar C_{\alpha'}) 
+\delta Z(\bar C_\beta) +\delta Z(\bar C_{\beta'}) \\
\quad
+\delta Z_{11} +\delta Z_{12} +\delta\tilde W_1 
  \end{array}
\right] \,.
\nonumber 
\end{eqnarray}
Using the fact that up to $o(g_R^2)$ the two quantities $W_1$ and
$\tilde W_1$ are essentially the same\footnote{Only a factor
  $\mu^{D-4}$ comes in since $g_R$ as given in (\ref{grug}) is
  dimensionless in contrast to $g$.} we find
\begin{eqnarray}
\lefteqn{\lim_{D\to 4} \delta^{ab} {2N\over N^2 -1} \mu^{D-4}
\lim_{x'\to x \atop y'\to y}\partial_\mu^x \partial_\nu^y \,
\delta\tilde W_1}
\nonumber\\ &&
= \lim_{D\to 4} \delta^{ab} {2N\over N^2 -1} 
{1\over (ig)^2}\lim_{x'\to x \atop y'\to y}\partial_\mu^x 
\partial_\nu^y \,
\left(1+(ig)^2 \delta W_1 \right)
\nonumber\\ &&
= \lim_{D\to 4} \delta^{ab} {2N\over N^2 -1} 
{1\over (ig)^2}\lim_{x'\to x \atop y'\to y}\partial_\mu^x 
\partial_\nu^y \,
W_1 \,\Big\vert_{g=0}
\nonumber\\ &&
=\lim_{D\to 4}
\langle A^a_\mu(x) A^b_\nu(y) \rangle^0_{\mbox{\scriptsize rad}} \,.
\label{naipa}\end{eqnarray}
To get the $\delta Z$'s we must calculate $\delta\tilde W_1$ which is 
straightforward using (\ref{w1def}) and (\ref{roper}). 
We only need the Feynman propagator (\ref{feynman}) to get
\begin{eqnarray}
\lefteqn{\delta\tilde W_1 = 
-\mu^{4-D} {N^2 -1\over 2N} {\Gamma(D/2-1) \over 4\pi^{D/2}}} 
\nonumber\\ &&
\left[
\left(
\vert x' \vert^{4-D} +\vert x-x' \vert^{4-D} + \vert x \vert^{4-D} + 
\vert y' \vert^{4-D} +\vert y-y' \vert^{4-D} + \vert y \vert^{4-D}
\right) I_1 
\right.\nonumber\\ && 
{}+ I_2(x',x-x') + I_2(x-x',-x) + I_2(-x,x')
\nonumber \\ &&
{}+ I_2(y',y-y') + I_2(y-y',-y) + I_2(-y,y')
\nonumber \\ &&
{}- I_2(x',-y') + I_2(x',-y) + I_2(y',-x) - I_2(x,-y)
\nonumber \\ && 
{}- I_3(y',-x',y-y') + I_3(y',-x,y-y') - I_3(x'-y',x-x',y'-y)
\nonumber \\ && 
{}- I_3(x',-y',x-x') + I_3(x',-y,x-x')
\Big] \label{dw1l}
\end{eqnarray}
with 
\begin{equation}
I_1:= \int\limits_0^1 \!\! ds  \int\limits_0^1 \!\! dt \,\Theta(s-t) \,
{1 \over [(s-t)^2]^{D/2-1}} \,,
\end{equation}
\begin{equation}
I_2(p,q):= \int\limits_0^1 \!\! ds  \int\limits_0^1 \!\! dt \,
{p\cdot q \over [(sp+tq)^2]^{D/2-1}}  \,,
\end{equation}
and
\begin{equation}
I_3(m,p,q):= \int\limits_0^1 \!\! ds  \int\limits_0^1 \!\! dt \,
{p\cdot q \over [(m+sp+tq)^2]^{D/2-1}}  \,.
\end{equation}
In the following we are interested only in the divergent parts of
these integrals. The integrals $I_1$ and $I_2$ are calculated in
Appendix~\ref{siint}. The results are
\begin{equation}
I_1 = -{1\over 4-D} + \mbox{finite} \label{inti1}
\end{equation}
and
\begin{equation}
  I_2(p,q) = {1\over 4-D} \, \gamma \cot\gamma + \mbox{finite}
  \label{inti2}
\end{equation}
where $\gamma$ is the angle between $p$ and $q$. The integral $I_3$ is
finite as long as $m\neq 0$.

To specify the renormalization factors $Z$ we choose the minimal
subtraction scheme $K^{\mbox{\scriptsize MS}}_\gamma$ as described in
\cite{korrad}. In dimensional regularization all the divergences are
given by sums of pole terms.  We define every $Z$ factor to be given
just by the respective sum. The important property of this
renormalization scheme is that the $Z$ factors depend on the angles
only and not on the length of the loop or of any part of the loop.
Using (\ref{inti1} and (\ref{inti2})) the $Z$ factors can be read off
from (\ref{dw1l}) (c.f.~\cite{brandt2}):
\begin{eqnarray}
\delta Z(\bar C_\alpha) 
&=& 
{N^2 -1\over 2N}{1\over 4\pi^2} {1\over 4-D} \, 
(\alpha \cot\alpha -1) \,,
\\
\delta Z(\bar C_{\alpha'}) 
&=& 
{N^2 -1\over 2N}{1\over 4\pi^2} {1\over 4-D} \, 
(\alpha' \cot\alpha' -1) \,,
\\
\delta Z(\bar C_\beta) 
&=& 
{N^2 -1\over 2N}{1\over 4\pi^2} {1\over 4-D} \, 
(\beta \cot\beta -1) \,,
\\
\delta Z(\bar C_{\beta'})
&=& 
{N^2 -1\over 2N}{1\over 4\pi^2} {1\over 4-D} \, 
(\beta' \cot\beta' -1) \,,
\\
\delta Z_{11} 
&=& 
{N^2 -1\over 2N}{1\over 4\pi^2} {1\over 4-D} 
\left[
(\gamma_{x'y} \cot\gamma_{x'y} -1) + 
           (\gamma_{xy'} \cot\gamma_{xy'} -1) 
\right] \,, \nonumber \\ &&
\\
\delta Z_{12} 
&=& 
{N^2 -1\over 2N}{1\over 4\pi^2}  {1\over 4-D} 
\left[
\gamma_{xx'} \cot\gamma_{xx'}  + \gamma_{yy'} \cot\gamma_{yy'} 
\right.\nonumber\\ && %\phantom{MM} 
\left.
{}- (\pi-\gamma_{x'y'})\cot(\pi-\gamma_{x'y'})
- (\pi-\gamma_{xy})\cot(\pi-\gamma_{xy})
\right]  \,.
\end{eqnarray}
Now we exploit the fact that only one of the angles, namely $\gamma_{xy}$,
depends on $x$ {\it and} $y$. All the other ones depend only on $x$ or $y$ 
separately, or on none of them. This simplifies (\ref{freepr}) drastically:
\begin{eqnarray}
\lefteqn{\langle A^a_\mu(x) A^b_\nu(y) \rangle_R^0 = 
\lim_{D\to 4} \delta^{ab} {2N\over N^2 -1} \mu^{D-4}
\lim_{x'\to x \atop y'\to y}\partial_\mu^x \partial_\nu^y \,
\left( \delta Z_{12} +\delta\tilde W_1 \right)}
\\ &&
=\lim_{D\to 4} \left(
\delta^{ab} \mu^{D-4} \partial_\mu^x \partial_\nu^y 
\left(
{1\over 4\pi^2} {1\over 4-D} (\pi-\gamma_{xy})\cot\gamma_{xy} 
\right)
+\langle A^a_\mu(x) A^b_\nu(y) \rangle^0_{\mbox{\scriptsize rad}}
\right)
\nonumber\end{eqnarray}
where we have used (\ref{naipa}) to get the last expression.

%%%%%%%%%%%%%%%%%%%%5 Appendix 3 %%%%%%%%%%%%%%%%%%%%%%%%%%%%%%%%%%%%%
\section{Some Important Integrals} \label{siint}

The integrals $I_1$ and $I_2$ played an important part in the
renormalization procedure of appendix~\ref{app:renprog} and determine
the divergences of the naive free radial propagator introduced in
section~\ref{sec:radprop}. They are discussed in detail below.

To calculate
\begin{equation}
I_1:= \int\limits_0^1 \!\! ds  \int\limits_0^1 \!\! dt \,\Theta(s-t) \,
{1 \over [(s-t)^2]^{D/2-1}} 
\end{equation}
we introduce the substitution
\begin{equation}
g=s-t \quad,\quad h=s+t  
\end{equation}
to get
\begin{eqnarray}
I_1 &=& 
{1\over 2} \int\limits_0^1 \!\! dg  \int\limits_g^{2-g} \!\! dh \, g^{2-D}
= \int\limits_0^1 \!\! dg \, (1-g) \, g^{2-D} \nonumber\\
&=& {\Gamma(2) \Gamma(3-D) \over \Gamma(5-D)} = {1\over (4-D)(3-D)} \,.
\label{i1erg}\end{eqnarray}

For the calculation of 
\begin{equation}
I_2(p,q):= \int\limits_0^1 \!\! ds  \int\limits_0^1 \!\! dt \,
{p\cdot q \over [(sp+tq)^2]^{D/2-1}}  \,. \label{appi2}
\end{equation}
we have to distinguish the two cases $p \neq  \alpha q$ where the only
divergence that appears is for $s=t=0$ and $p = \alpha q$ with an additional
divergence at $s=t\alpha$. Here we will only need the former.

\begin{figure}
\begin{center}
\begin{minipage}{10cm}
\epsfysize=4cm
\epsfbox{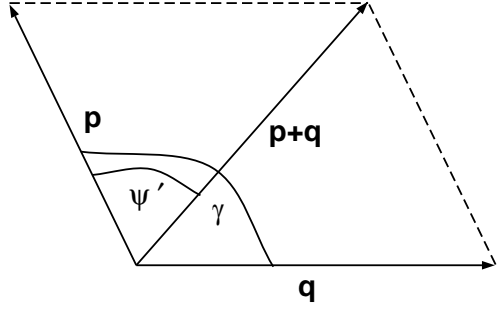}
\end{minipage}
\caption{The geometry of the variables appearing in the 
  calculation of $I_2$
  (\ref{appi2}).} \label{fig:pqangl}
\end{center}
\end{figure}

As a first step it is useful to separate off the divergence at the
origin by the substitution
\begin{equation}
\lambda = s+t \quad,\quad x=s/\lambda  \,.
\end{equation}
This yields
\begin{eqnarray}
  I_2(p,q) &=& \left( \int\limits_0^{1/2} \!\! dx
    \int\limits_0^{1/(1-x)} \!\!\!\! d\lambda \, + \int\limits_{1/2}^1
    \!\! dx \int\limits_0^{1/x} \!\! d\lambda \right) \lambda^{3-D}
  {p\cdot q \over [(xp+(1-x)q)^2]^{D/2-1}} \nonumber\\ &=&
  \int\limits_0^{1/2} \!\! dx \,{(1-x)^{D-4} \over 4-D} \, {p\cdot q
    \over [(xp+(1-x)q)^2]^{D/2-1}} \nonumber\\ && +\int\limits_{1/2}^1
  \!\! dx \,{x^{D-4} \over 4-D} \, {p\cdot q \over
    [(xp+(1-x)q)^2]^{D/2-1}} \label{xint}\,.
\end{eqnarray} 
As long as $p \neq -q$ holds there are no divergences in the
$x$-integration since
\begin{equation}
u(x):=xp+(1-x)q
\end{equation}
never vanishes. We introduce the angle between $p$ and $q$
\begin{equation}
\cos\gamma := {p\cdot q \over \vert p \vert \, \vert q\vert }
\end{equation}
and the substitution \cite{korrad}
\begin{equation}
e^{2i\psi} = {x\vert p\vert +(1-x)\vert q\vert e^{i\gamma} \over
x\vert p\vert +(1-x)\vert q\vert e^{-i\gamma}} \label{psisub}\,.
\end{equation}
Note that $\psi$ is nothing but the angle between $p$ and $u(x)$. 
To perform this substitution in (\ref{xint}) we need
\begin{equation}
x = \vert q\vert \sin(\gamma-\psi) /N(\psi) \quad,\quad
1-x = \vert p \vert \sin\psi / N(\psi)  \,,
\end{equation}
\begin{equation}
[u(x)]^2 = p^2 q^2 \sin^2\gamma /[N(\psi)]^2 
\quad\mbox{and} \quad
{d\psi \over dx} = 
- {[N(\psi)]^2 \over \vert p\vert\,\vert q\vert \sin\gamma } 
\end{equation}
with
\begin{equation}
  N(\psi) := \vert p \vert \sin\psi + \vert q\vert \sin(\gamma-\psi)
  \,.
\end{equation}
In addition it is useful to introduce
\begin{equation}
\psi' :=\psi(x=1/2)
\end{equation}
which is the angle between $p$ and $p+q$ (cf.~Fig.~\ref{fig:pqangl}).

Using all that we end up with
\begin{eqnarray}
\lefteqn{
I_2(p,q) = \int\limits_\gamma^{\psi'}\!\! d\psi \, 
{-\vert p\vert\,\vert q\vert \sin\gamma  \over N^2}  
\left({\vert p \vert \sin\psi \over N}\right)^{D-4}
\left({N^2 \over p^2 q^2 \sin^2\gamma} \right)^{D/2-1} {p\cdot q \over 4-D}}
&&\nonumber\\
&& +\int\limits_{\psi'}^0\!\! d\psi \, 
{-\vert p\vert\,\vert q\vert \sin\gamma  \over N^2}  
\left({ \vert q\vert \sin(\gamma-\psi) \over N}\right)^{D-4}
\left({N^2 \over p^2 q^2 \sin^2\gamma} \right)^{D/2-1} {p\cdot q \over 4-D}
\nonumber \\
  &=& {- \cos\gamma \sin^{3-D}\gamma\over 4-D} \left( \vert
    q\vert^{4-D} \int\limits_\gamma^{\psi'}\!\! d\psi \,
    \sin^{D-4}\psi + \vert p\vert^{4-D} \int\limits_{\psi'}^0\!\!
    d\psi \, \sin^{D-4}(\gamma-\psi) \right) \nonumber\\ 
&=& {- \cos\gamma \sin^{3-D}\gamma \over
    4-D} \left( \vert q\vert^{4-D}
    \int\limits_\gamma^{\psi'}\!\! d\psi \, \sin^{D-4}\psi + \vert
    p\vert^{4-D} \int\limits_\gamma^{\gamma-\psi'}\!\! d\psi \,
    \sin^{D-4}\psi \right) \nonumber\\ &=& {1\over 4-D} \,\gamma
  \cot\gamma \; +\mbox{finite}
 \label{i2neq}\,. 
\end{eqnarray}
\end{appendix}

%%%%%%%%%%%%%%%%% Bibliography  %%%%%%%%%%%%%%%%%%%%%%%%%%%%%%%%%%%%%%


\begin{thebibliography}{}
\bibitem{pascual} P.~Pascual and R.~Tarrach, ``QCD: Renormalization for the
Practitioner'', Lecture Notes in Physics, Vol. 194 (Springer, Berlin, 1984). 
\bibitem{cheng} H.~Cheng and E.-C.~Tsai, Phys.~Rev.~Lett. {\bf 57} (1986) 511.
\bibitem{cara} S.~Caracciolo, G.~Curci, and P.~Menotti, Phys.~Lett. {\bf B113}
(1982) 311.
\bibitem{leroy} J.-P.~Leroy, J.~Micheli, and G.-C.~Rossi, Z.~Phys. {\bf C36}
(1987) 305.
\bibitem{christ} N.H.~Christ and T.D.~Lee, Phys.~Rev. {\bf D22} (1980) 939.
\bibitem{doust} P.~Doust, Ann.~Phys. {\bf 177} (1987) 169.
\bibitem{taylor} P.J.~Doust and J.C.~Taylor, Phys.~Lett. {\bf B197} (1987) 232.
\bibitem{shif} M.A.~Shifman, Nucl.~Phys. {\bf B173} (1980) 13.
\bibitem{fock} V.A.~Fock, Sov.~Phys. {\bf 12} (1937) 404.
\bibitem{schwing} J.~Schwinger, Phys.~Rev. {\bf 82} (1952) 684.
\bibitem{cron} C.~Cronstr\"om, Phys.~Lett. {\bf B90} (1980) 267.
\bibitem{kumm} W.~Kummer and J.~Weiser, Z.~Phys. {\bf C31} (1986) 105.
\bibitem{moda}   G.~Modanese, J.~Math.~Phys. {\bf 33} (1992) 1523.
\bibitem{menot} P.~Menotti, G.~Modanese, and D.~Seminara, Ann.~Phys. {\bf 224}
(1993) 110.
\bibitem{azam} M.~Azam, Phys.~Lett. {\bf B101} (1981) 401.
\bibitem{bralic} N.E.~Bralic, Phys.~Rev. {\bf D22} (1980) 3090.
\bibitem{ElGyuVa86} H.-Th.~Elze, M.~Gyulassy, and D. Vasak, 
Nucl. Phys. {\bf B276} (1986) 706.
\bibitem{korrad} G.P.~Korchemsky and A.V.~Radyushkin, Nucl.~Phys. {\bf B283} 
(1987) 342.
\bibitem{regul} V.S.~Dotsenko and S.N.~Vergeles, Nucl.~Phys. {\bf B169}
(1980) 527.
\bibitem{brandt} R.A.~Brandt, F.~Neri, and M.-A.~Sato, Phys.~Rev. {\bf D24} 
(1981) 879.
\bibitem{brandt2} R.A.~Brandt, A.~Gocksch, M.-A.~Sato, and F.~Neri, 
Phys.~Rev. {\bf D26} (1982) 3611.
\bibitem{korkor} I.A.~Korchemskaya and G.P.~Korchemsky, Phys.~Lett. {\bf B287}
(1992) 169.
\end{thebibliography}
\end{document}